# CADgpt: Harnessing Natural Language Processing for 3D Modelling to Enhance Computer-Aided Design Workflows

Authors
**Timo Kapsalis**[a] – t.kapsalis@derby.ac.uk
[a]: University of Derby, UK

## Abstract

This paper introduces *CADgpt*, an innovative plugin integrating Natural Language Processing (NLP) with Rhino3D for enhancing 3D modelling in computer-aided design (CAD) environments. Leveraging OpenAI's GPT-4, CADgpt simplifies the CAD interface, enabling users, particularly beginners, to perform complex 3D modelling tasks through intuitive natural language commands. This approach significantly reduces the learning curve associated with traditional CAD software, fostering a more inclusive and engaging educational environment. The paper discusses CADgpt's technical architecture, including its integration within Rhino3D and the adaptation of GPT-4 capabilities for CAD tasks. It presents case studies demonstrating CADgpt's efficacy in various design scenarios, highlighting its potential to democratise design education by making sophisticated design tools accessible to a broader range of students. The discussion further explores CADgpt's implications for pedagogy and curriculum development, emphasising its role in enhancing creative exploration and conceptual thinking in design education.

**Keywords**: Natural Language Processing, Computer-Aided Design, 3D Modelling, Design Automation, Design Education, Architectural Education

## Introduction

In the evolving narrative of Computer-Aided Design (CAD), the interface between human creativity and computational precision has been a subject of continuous refinement [1,2]. The trajectory from command-line to graphical user interfaces in CAD software marked a significant leap in usability and accessibility [3]. Yet, the integration of Artificial Intelligence (AI), and specifically natural language processing (NLP), promises a further democratisation of design tools, enabling a more intuitive interaction that aligns with human cognitive processes [4,5]. We introduce CADgpt, a groundbreaking plugin for Rhino3D that leverages OpenAI's GPT-4 model, to extend the frontier of design education and practice.

The development of CADgpt is underpinned by a critical recognition of the challenges faced by nascent design students when confronted with the complexity of traditional CAD interfaces [6]. Previous evidence from educational settings reveals a persistent barrier: the steep learning curve that can stifle the creative exploration of students new to 3D modelling environments [6,7]. CADgpt aims to dismantle this barrier by allowing users to employ natural language to execute complex modelling tasks, thereby facilitating a more inclusive and engaging learning experience.

The purpose of CADgpt transcends mere technical innovation; it is a pedagogical tool fashioned to integrate seamlessly into the design curriculum. By enabling command execution through conversational language, CADgpt aligns with cognitive learning theories which posit that language is a critical mediator in the construction of knowledge. This alignment suggests that by reducing the technical overhead of CAD software, students can focus more on the creative aspects of design, fostering a more profound educational experience.

The potential impact of CADgpt on the field of design education is profound. By abstracting the technical complexities of 3D modelling into intuitive language, CADgpt stands to broaden the accessibility of design education, enhance the efficiency of learning, and enable a wider demographic of students to engage with advanced design tools. This shift not only empowers students of diverse backgrounds and abilities but also invites a reassessment of the role of AI in the creative process, suggesting new paradigms of interaction between designers and their tools.

In the following sections, this paper will detail the technical architecture of CADgpt, illustrate its application through examples from educational settings, and discuss the broader implications for design educators, students, and the future of design education in an era increasingly characterised by the symbiosis of human creativity and computational intelligence.

**Technical description**

CADgpt represents an interactive integration of OpenAI's GPT-4 model [8] with Rhino3D CAD environment [9]. This section delineates the technical constitution of CADgpt, elucidating the operational mechanics that empower users to engage with Rhino3D through the medium of natural language.

CADgpt is architected as a plugin that functions within the Rhino3D software ecosystem. It serves as an intermediary, translating natural language inputs into the Grasshopper scripting and command framework. This translation is facilitated through a custom API that leverages the GPT-4 model's advanced NLP capabilities. The API acts as a conduit, parsing user inputs, interpreting their design intentions, and generating corresponding RhinoScript or Grasshopper definitions that Rhino3D can execute.

At the heart of CADgpt is the GPT-4 AI model, a state-of-the-art language processor renowned for its ability to understand and generate human-like text [8]. The engine processes input commands in natural language, comprehends the context and intent, and translates these into a structured command set that is compatible with CAD operations. This process includes an understanding of geometric concepts, design terminology, and procedural instructions, all within the vernacular of design automation.

A typical interaction with CADgpt begins with the user articulating a design task in natural language. This input is processed by GPT-4, which generates a script that includes necessary variables, functions, and logic. The output script is then passed to Rhino3D's command processor. For example, a user may input "create a 3D model of a ten-story building with a glass facade", which CADgpt will interpret and translate into a series of Rhino3D actions, creating the geometric model as specified. Figure 1 presents a workflow with basic operations of the CADgpt tool.

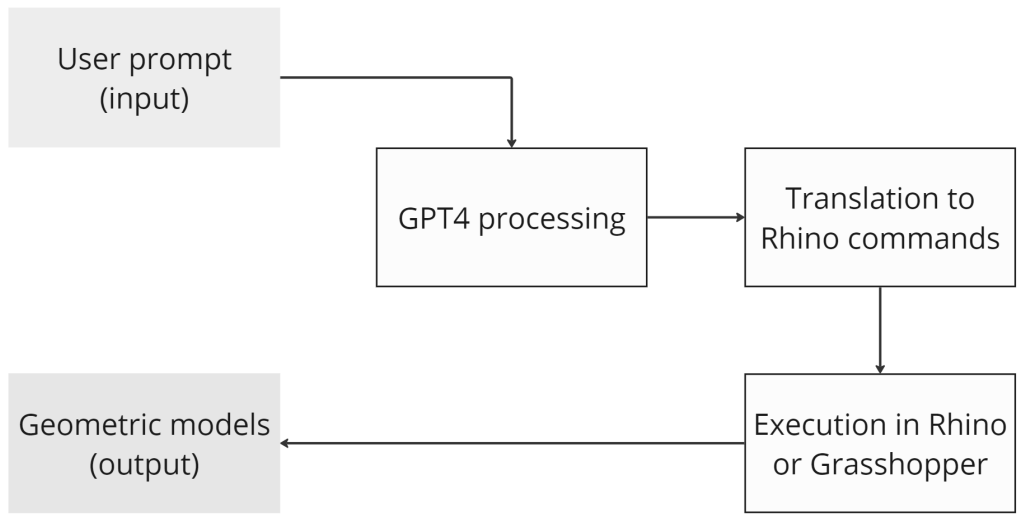

*Figure 1: The CADgpt operational workflow*

CADgpt employs custom algorithms (e.g., Natural Language Understanding and Geometric Modelling algorithms) to refine the output from the GPT-4 model, ensuring the generated commands are optimised for Rhino3D's environment. These algorithms address the specificities of CAD tasks, such as the precision required in geometric calculations and the hierarchical structuring of design elements. The optimisation process also includes error checking and feedback loops, which allow CADgpt to learn from interactions and improve over time.

The CADgpt plugin is designed with a minimalist interface to maintain focus on the Rhino3D environment while providing clear and concise feedback to the user. Visual cues and command confirmation are built into the system to guide users through the design process and ensure clarity in command execution.

In the construction of CADgpt, careful attention has been paid to the ethical implications of AI integration in design tools. All interactions with the GPT-4 model are conducted in accordance with OpenAI's use-case policies, ensuring data privacy and security [8]. Additionally, the plugin includes safeguards to prevent the generation of designs that could be ethically problematic or unsafe.

The technical architecture of CADgpt is a testament to the harmonious potential between human-centric design processes and AI efficiency. By enabling natural language as a means to command and control within a CAD environment, CADgpt stands as a significant advance in the field of design automation, poised to redefine the interface between designers and digital creation tools.

**Early examples of CADgpt implementation**

The practical applications of CADgpt are best illustrated through case studies that highlight its functionality and user experience. Here we present three such examples, showcasing the tool's use in varying contexts. All examples refer to student exercises within the Architectural Technology program at the University of Derby.

*Example 1: geometric interplay*

The attached image demonstrates a fundamental operation where a student used CADgpt to create a complex geometric model. The task was to intersect a 100x100x30 cm box with a sphere of 30 cm radius, positioned at a random edge, and then execute a union of the two shapes within Rhino3D. The student simply input the natural language command: "Create a 100x100x30 cm box, which is intersected by a sphere of 30 cm radius at a random edge. Bake their union on Rhino". CADgpt interpreted this command, generated the corresponding geometric shapes, and performed the Boolean union operation—all without the student needing to manually input any traditional Rhino3D commands or navigate complex toolbars (Figure 2).

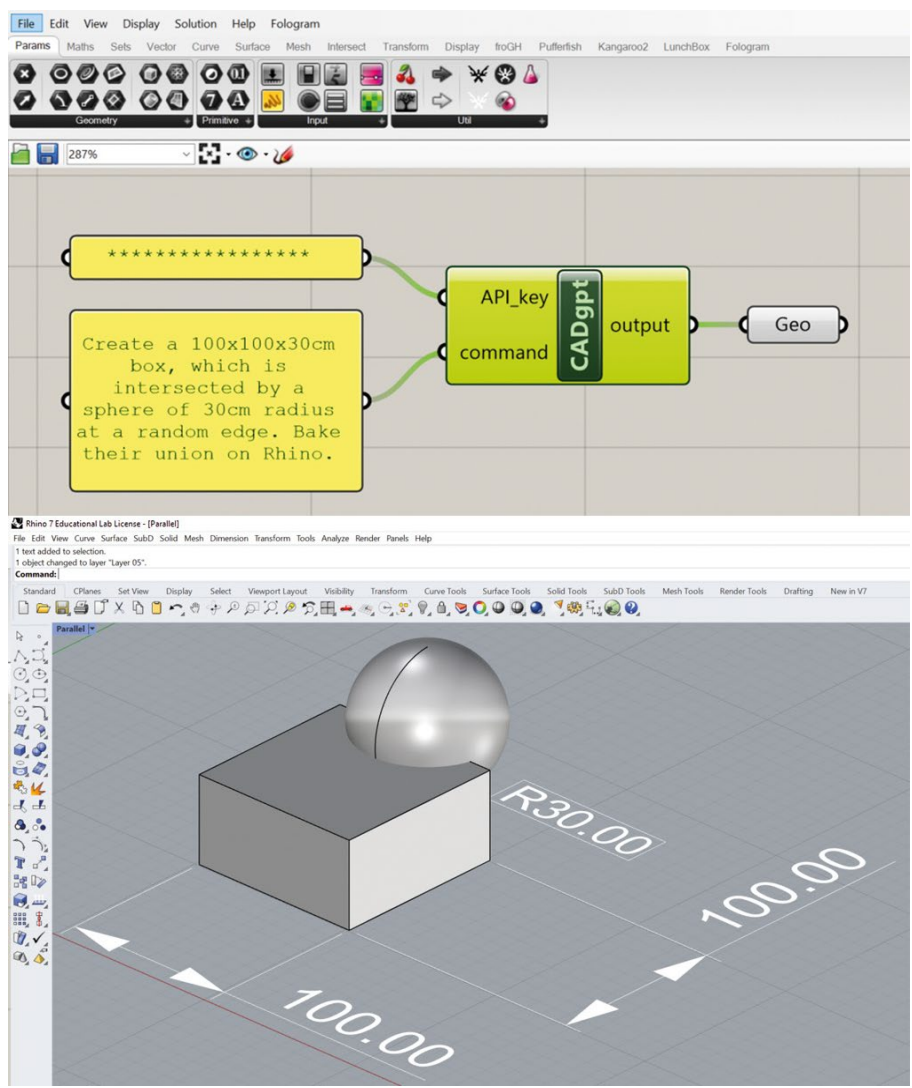

*Figure 2: Using CADgpt for simple geometric modelling, an example.*

*Example 2: architectural prototyping*

In an architectural design workshop, students were tasked with conceptualising a pavilion with organic forms, taking inspiration from precedent projects. Using CADgpt, one student described their vision: "Design a pavilion with a hyperbolic canopy, inspired by the Candela structures". CADgpt processed this description and provided a step-by-step Grasshopper definition, which iteratively refined the pavilion's form based on the student's further natural language feedback, allowing for a rapid prototyping process that typically would require advanced Rhino3D proficiency (Figure 3).

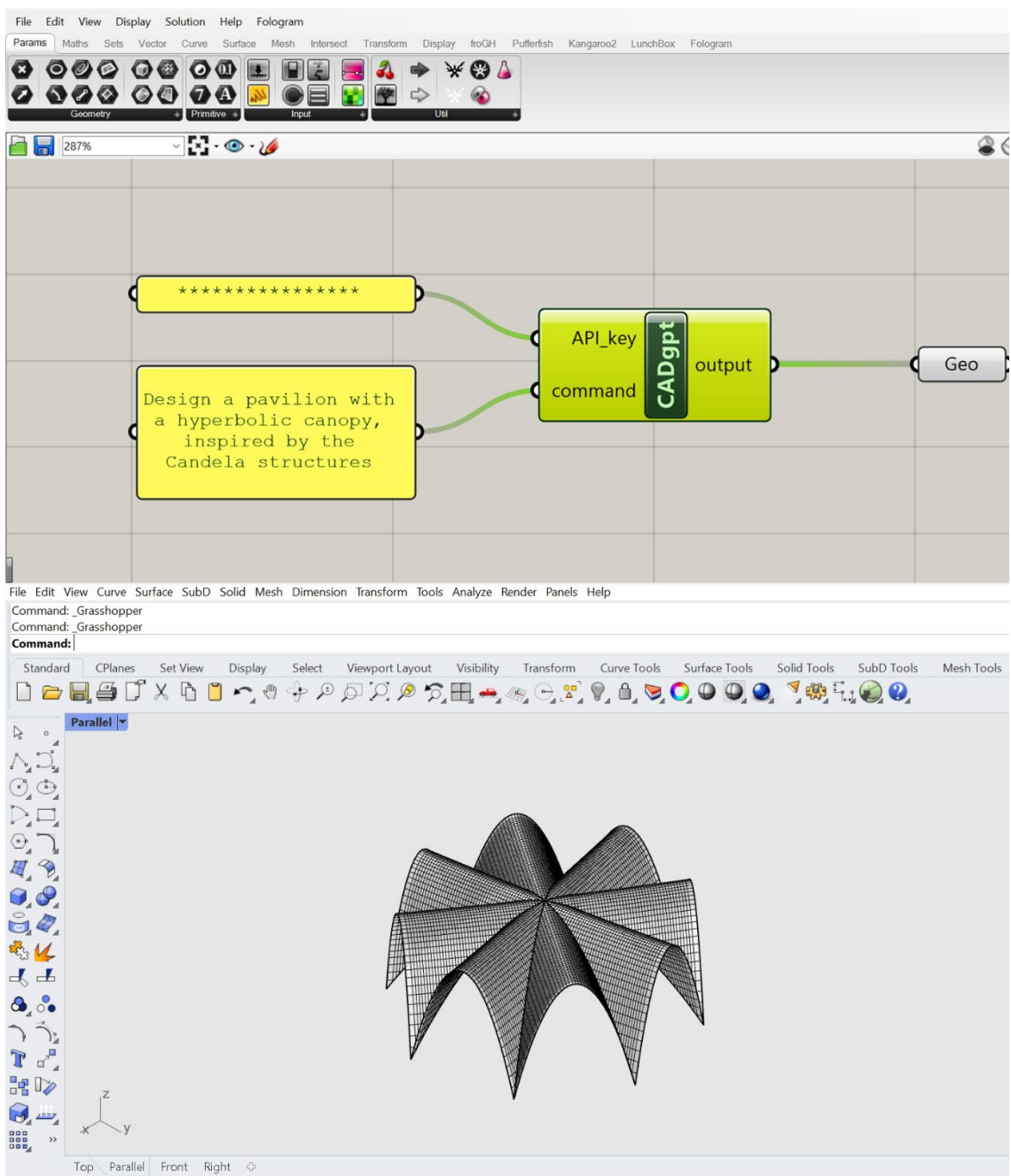

*Figure 3: Using CADgpt for early-stage architectural prototyping, an example.*

*Example 3: urban design simulation*

During a course on sustainable urban design, students used CADgpt to simulate sunlight patterns across a planned residential area. A student inputted: "Generate a grid of buildings 15 meters high, spaced 20 meters apart, and simulate the sunlight paths during the UK summer solstice". CADgpt recognised the command, created the grid of buildings, and utilised Rhino3D's sun path simulation tools to display the sunlight patterns (Figure 4). This allowed the student to conduct shade studies to assess the potential for passive solar heating and make informed decisions about building placement and design for energy efficiency.

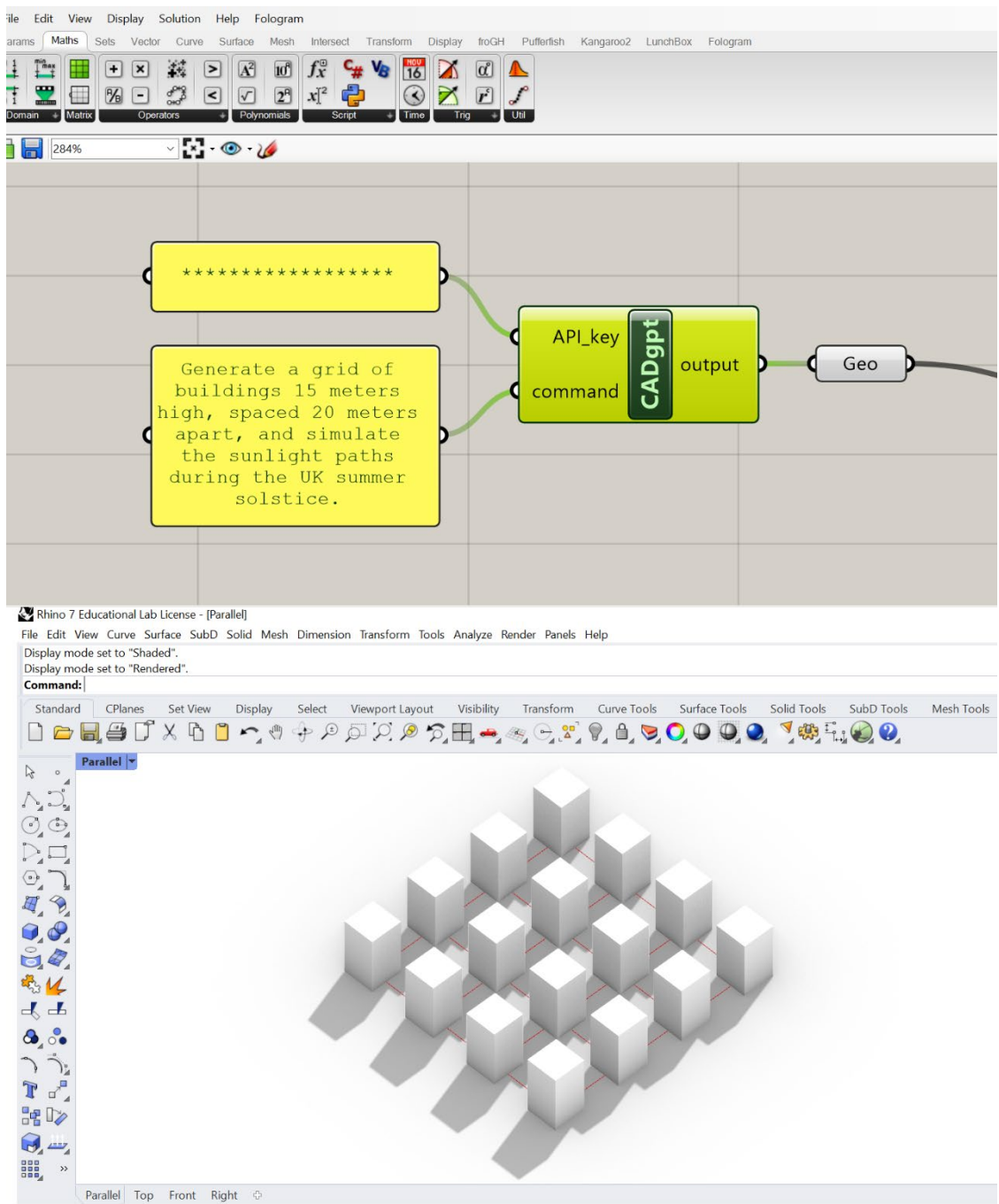

*Figure 4: Using CADgpt for urban design tasks, an example.*

These examples underscore the versatility of CADgpt in translating natural language into complex CAD operations, enabling students to engage with advanced design scenarios without the prerequisite of in-depth CAD knowledge. Each case study reflects the tool's potential to expand the boundaries of design education, making sophisticated design and analysis tasks accessible to a broader range of students.

**Discussion**

As CADgpt remains in a work-in-progress and trial stage, its introduction into the design curriculum is a pioneering step that necessitates both cautious optimism and critical reflection. This emergent tool offers an opportunity to fundamentally reassess the pedagogical strategies employed in design education, with broad implications for educators, students, and the wider field of design automation research.

CADgpt embodies the pedagogical principle that tool transparency can enhance learning outcomes. By reducing the cognitive load associated with learning complex software commands, educators can refocus their pedagogy on design thinking and problem-solving [10,11]. CADgpt aligns with the constructivist approach to education, which emphasises the importance of active, contextualised learning processes where students construct new knowledge based on their experiences [12,13].

The implementation of CADgpt necessitates a revaluation of design curriculum structures. Courses traditionally structured around the gradual introduction of software functionality may need to be adapted to integrate AI tools like CADgpt, which change the way students interact with CAD software. This integration also opens the door to a more interdisciplinary curriculum that can include students from various backgrounds, promoting diversity within the field of design.

One of the most compelling contributions of CADgpt is its potential to liberate student creativity from the technical constraints of traditional CAD learning [14,15]. By simplifying interaction with CAD software through natural language processing, students can allocate more time and cognitive resources to conceptual development and creative experimentation. As early results indicated, CADgpt facilitates a shift away from the intricacies of command syntax and workflow memorisation, allowing students to engage with the essence of design itself - the realisation of innovative concepts and solutions.

For design students, especially those in the nascent stages of their education, CADgpt offers a streamlined path to developing CAD proficiency. By interacting with the software through natural language, students can engage with 3D modelling without the intimidation of intricate toolsets [16]. It is therefore possible that this accessibility can accelerate the acquisition of spatial reasoning and design skills, which are fundamental to the architectural and design disciplines.

By lowering language and technical barriers, CADgpt contributes significantly to inclusive education and adaptive learning. By providing a platform that responds to natural language commands, CADgpt inherently adapts to the diverse cognitive and linguistic styles of its users. This adaptability ensures that students can engage with the tool in a manner that aligns with their personal learning preferences, thus supporting a more personalised educational journey. The adaptive nature of CADgpt is particularly salient in its potential to cater to students with disabilities or those who might otherwise find the technical rigor of traditional CAD interfaces a barrier. The tool's capacity to learn and adjust to individual student interactions embodies the principles of adaptive learning—where technology is used to tailor educational experiences to each student's needs and abilities [17,18]. As such, the tool has the potential to democratise design education, making it more accessible to a wider range of learning styles and abilities.

CADgpt can serve as a spark for research into the intersection of AI and design. It provides a practical example of how AI can be integrated into design tools, inviting further investigation into how such integration can evolve. Research can explore the implications of AI-assisted design on creativity, the development of new design methodologies, and the future role of designers in an AI-augmented design process. Looking beyond education to professional practice, CADgpt suggests a future where designers may collaborate more closely with AI, utilising natural language to communicate with CAD systems. This could lead to more intuitive design processes and potentially change the role of designers, who may increasingly become orchestrators of AI-assisted design rather than sole creators [19-21].

As CADgpt evolves, its development will be directed towards refining its ability to understand and execute increasingly complex and abstract design requests. This progression will entail training the AI with diverse datasets that capture the breadth of design terminology and conceptual descriptions. The aim is to develop a tool that not only understands detailed commands but can also engage with the ambiguous and often non-linear nature of creative design processes. Further development of CADgpt will focus on expanding its capabilities to interpret more complex design tasks and integrating with additional specialised plugins, enhancing its versatility.

CADgpt represents a transformative tool that has the potential to reshape the landscape of design education and practice. By enabling natural language interaction with CAD software, it simplifies the learning process, fosters inclusivity, and opens up new avenues for research and practice in design automation. As the tool evolves and its usage becomes more widespread, it is likely to have a lasting impact on how we teach, learn, and practice design in an increasingly digital world.